\documentclass[pre,aps,twocolumn,showpacs]{revtex4}

\usepackage{dcolumn}
\usepackage{amssymb}
\usepackage{amsmath}
\usepackage{graphicx}
\usepackage{epsf,epsfig}

\newcommand \be {\begin{equation}}
\newcommand \ee {\end{equation}}
\newcommand \bea {\begin{eqnarray}}
\newcommand \eea {\end{eqnarray}}
\newcommand \ve {\varepsilon}

\begin{document}

\title{Entropy-based characterizations of the observable-dependence of the fluctuation-dissipation temperature}

\author{Kirsten Martens$^{1,2}$, Eric Bertin$^3$, and Michel Droz$^2$}
\affiliation{
$^1$ Universit\'e de Lyon; Universit\'e Lyon 1, Laboratoire de Physique
de la Mati\`ere Condens\'ee et des Nanostructures; CNRS, UMR 5586,
43 Boulevard du 11 Novembre 1918, F-69622 Villeurbanne Cedex, France\\
$^2$ Department of Theoretical Physics, University of Geneva,
CH-1211 Geneva 4, Switzerland\\
$^3$ Universit\'e de Lyon, Laboratoire de Physique, ENS Lyon, CNRS,
46 All\'ee d'Italie, F-69007 Lyon, France}
\date{\today}

\begin{abstract}
The definition of a nonequilibrium temperature through generalized
fluctuation-dissipation relations relies on the independence of the
fluctuation-dissipation temperature from the observable
considered. We argue that this observable independence is deeply related
to the uniformity of the phase-space probability distribution on the
hypersurfaces of constant energy. This property is shown explicitly on
three different stochastic models, where observable-dependence of the
fluctuation-dissipation temperature arises only when the uniformity of
the phase-space distribution is broken.
The first model is an energy transport model on a ring,
with biased local transfer rules.
In the second model, defined on a fully connected geometry,
energy is exchanged with two heat baths at different
temperatures, breaking the uniformity of the phase-space distribution.
Finally, in the last model,
the system is connected to a zero temperature reservoir, and
preserves the uniformity of the phase-space distribution
in the relaxation regime, leading to an observable-independent temperature.
\end{abstract}

\pacs{05.70.Ln, 05.10.Cc, 05.20.-y}

\maketitle

\section{Introduction}

The definition of macroscopic quantities that can characterize
nonequilibrium systems is a challenging and active field in statistical
physics. Several approaches have been proposed in the literature,
by generalizing thermodynamic \cite{Jou} or statistical physics
approaches \cite{Edwards,Crisanti,Cugl-Kurchan,Hatano,Ritort}.
One of these approaches is based on the possible generalization
to nonequilibrium situations of intensive thermodynamic parameters
(chemical potential, compactivity...) conjugated to conserved
quantities \cite{Barrat00,itp-short, itp-long}.
Such parameters are then defined as the logarithmic derivative
of a generalized partition function with respect to the
corresponding conserved quantity.
Yet, energy is in general not conserved in nonequilibrium systems,
and other approaches are necessary in order to define a nonequilibrium
temperature. Along this line, the introduction of effective temperatures
in nonequilibrium systems through generalized fluctuation-dissipation
relations (FDR) has played a major role \cite{CuKuPe,Crisanti}.

In equilibrium the FDR turns out to be a powerful tool to describe
the relaxation of slightly perturbed systems towards their equilibrium states.
The fluctuation-dissipation theorem relies on the principle that the
response of a system in thermodynamic equilibrium to a small perturbation
is the same as its response to a spontaneous fluctuation.
Accordingly, there is a direct relation between the fluctuation
characteristics of the thermodynamic system for a given observable
and its linear response.
This is a very strong property, because it is valid for all systems
in equilibrium, independent of the details of the microscopic dynamics
and the observable considered.
Hence the relation gives rise to a universal proportionality factor,
precisely given by the equilibrium temperature.

However, in a nonequilibrium system this relation is a priori not valid.
The theoretical investigation of the breaking of the fluctuation-dissipation
theorem in spin-glass systems \cite{CuKu93,CuKuPe}
and in turbulence \cite{Hohenberg}
has stimulated a wide range of experimental and numerical studies aiming
to define an effective temperature in many different systems,
ranging from granular materials \cite{Baldassarri,Makse,Danna,Puglisi}
to glasses \cite{Israeloff,Kob,Barrat-Berthier,Sciortino,Berthier-Barrat-A,Berthier-Barrat-B},
spin-glasses \cite{Ocio03,Ocio04},
gels and colloidal suspensions \cite{Ciliberto,Abou,Gomez},
liquid crystals \cite{Joubaud}, or turbulent flows \cite{Naert}.
Despite the large body of theoretical work devoted to nonequilibrium
generalizations of the FDR \cite{CuKuPe,Cugliandolo97,Franz,Kurchan00,Bertin-temp,Sasa,Kurchan05,Garriga,Levine06a,Levine06b,Seifert06,Mayer06,Corberi,Cugliandolo07,Levine08,Chetrite,Maes,Prost,Villamaina,Seifert09},
the question of the observable-dependence of the effective temperature
defined from the FDR has mainly been studied case by case in
specific models \cite{Berthier-Barrat-A,Berthier-Barrat-B,Baldassarri,Sollich02,Fielding03,Mayer03,Gambassi,Sollich05,Sollich06},
and no clear rationale has been proposed to interpret the
observable-dependence.

In this paper we study, following our recent letter \cite{fdrprl},
how the characteristics of a nonequilibrium
distribution of the microstates influence
the possibility to define an observable-independent temperature in the system.
We relate the observable-dependence of the FDR-temperature
to the ``lack of entropy'', defined as the entropy
difference between the nonequilibrium distribution and the equilibrium
distribution with the same energy.
We also observe that the entropy production,
which is a natural characterization
of nonequilibrium systems \cite{deGroot,Cugliandolo97,GallaCoh}, seems to
bear no systematic relation to the dependence of the temperature on the
choice of observable.
Our study is however restricted to systems that remain close
to an equilibrium state, and where correlation functions decay to zero
on a single time scale.
Most aging systems \cite{Bouchaud} are thus excluded from our analysis.

The paper is organized as follows. In Sect.~\ref{sec-frame} we present
the general framework, introducing a generalized fluctuation-dissipation
relation and the notion of lack of entropy. We also relate quantitatively
the observable-dependence of the effective temperature to
the lack of entropy.
In Sect.~\ref{sec-models}, this relation is quantitatively illustrated
on three different stochastic models.
First, an exactly solvable energy transport model on a ring in contact
to a reservoir is studied (Sect.~\ref{sec-ring}).
In this model, the internal flux results from the bulk dynamics rather
than from an external drive.
In the second example, an external drive is introduced (Sect.~\ref{sec-mean}).
More specifically, we consider a fully connected model in contact with two
heat baths at slightly different temperatures, resulting in a
nonequilibrium steady state \cite{fdrprl}.
As a last example, we discuss a variant of the latter model, in which the
system is connected to a single heat bath at zero temperature
(Sect.~\ref{sec-slow}). This dynamics leads to a slow relaxation towards
the ground state, during which the non-stationary distribution can be
computed.
Finally, Sect.~\ref{sec-discuss} summarizes and briefly discusses the
obtained results.


\section{General framework}
\label{sec-frame}

\subsection{Evaluation of the response function}

We first introduce the form of the generalized fluctuation-dissipation relation, that we use to define the out-of-equilibrium effective temperatures. We shall consider a generic system that is described by a set of $N$ variables $x_i$, $i=1,\ldots,N$. Since we are interested in the observable-dependence of the effective temperatures we introduce a family of observables $B_p$
indexed by an integer $p$.
In analogy to the equilibrium response theory, we are interested in the
dynamics of the observables due to the application of a perturbation to
a system that is in a nonequilibrium steady state, or relaxing to equilibrium.
This response will then be related to the fluctuations in the system
in the absence of perturbation.
In order to perturb the system,
a small external field $h$, conjugated to an observable $M$, can be
applied.
The following protocol allows for the definition of the linear response
of the observable $B_p$ to the external probe field.
The field $h$ takes a constant and small non-zero value until time $t_s$,
and it is then suddenly switched off.
The subsequent evolution of the observable $B_p$ then provides the
linear response to the probe field.
More precisely, the two-time linear response $\chi_p(t,t_s)$
is defined, for $t>t_s$, as
\begin{eqnarray}
 \chi_p(t,t_s)=\left.\frac{\partial}{\partial h}\right|_{h=0} \langle\!\langle
B_p(t,t_s) \rangle\!\rangle,
\end{eqnarray}
where $\langle\!\langle \cdots \rangle\!\rangle$ denotes an average
over the dynamics corresponding to the field protocol described above.

The basic idea of the FDR is to relate the linear response function
$\chi_p(t,t_s)$ to the correlation function (computed in the absence of field)
\be\label{def-corr}
C_p(t,t_s)=\langle (B_p(t)-\langle B_p(t)\rangle)\,(M(t_s)-\langle
M(t_s)\rangle)\rangle\;.
\ee
In general, such a relation is not necessarily linear. However, in cases when
it is linear, a FDR is said to hold, namely
\begin{equation} \label{FDR-p-ttw}
\chi_p(t,t_s) = \frac{1}{T_p(t_s)}\, C_p(t,t_s)\;, \qquad t>t_s \;.
\end{equation}
The proportionality factor is the inverse of the effective temperature $T_p$.
In equilibrium, $T_p$ depends neither on time nor on the observable,
it is simply the bath temperature. In contrast, out of equilibrium, $T_p$ can
be time-dependent, and it can a priori depend on $p$, that is,
on the observable.
In the specific case of nonequilibrium steady state,
the above FDR simplifies to, setting $t_s=0$,
\begin{equation} \label{FDR-p}
\chi_p(t) = \frac{1}{T_p}\, C_p(t)\;,
\end{equation}
where $T_p$ becomes time-independent.

In the following we will consider situations such that a
fluctuation-dissipation relation as given in
Eq.~(\ref{FDR-p-ttw}) or (\ref{FDR-p}) exists,
and we shall focus on the possible dependence of $T_p$ on the choice of
the observable $B_p$.
We shall mainly consider steady-state systems, but we will also
briefly study a nonstationary model (Sect.~\ref{sec-slow}), so that we
keep a time-dependent formalism.
The response of an observable to the perturbation can be formally rewritten using the distribution $\mathcal{P}(\{x_i\},h,t_s)$ of the microstate $\{x_i\} \equiv \{x_i,i=1,\ldots,N\}$ in the presence of the field $h$. To this aim we express $\langle\!\langle B_p(t,t_s)\rangle\!\rangle$ as
\begin{eqnarray} \label{obs}
\langle\!\langle B_p(t,t_s)\rangle\!\rangle &=& \int \prod_{i=1}^N dx_i dx_i'\, B_p(\{x_i\}) \times\nonumber\\
&&G_{t,t_s}^0(\{x_i\}|\{x_i'\}) \mathcal{P}(\{x_i'\},h,t_s) 
\end{eqnarray}
where $G_{t,t_s}^0(\{x_i\}|\{x_i'\})$ is the zero-field propagator,
that is the conditional probability to be in a microstate $\{x_i\}$ at time $t$ given that the system was in a microstate
$\{x_i'\}$ at time $t_s$, in the absence of the probe field.
Taking the derivative of Eq.~(\ref{obs}) with respect to $h$ at $h=0$,
and using the relation $\partial \mathcal{P}/\partial h = \mathcal{P}\partial \ln \mathcal{P}/\partial h$, we get
\be \label{response-rel-ttw}
\chi_p(t,t_s) = \left\langle B_p(t)\,
\frac{\partial \ln\mathcal{P}}{\partial h}(\{x_i(t_s)\},0,t_s)\right\rangle\;,
\ee
the average being computed at zero field.
Similar forms of this expression of the response function can be
found in the literature \cite{Agarwal,Villamaina,Prost,Seifert09}.
In the case of a steady-state system, with a distribution
$\mathcal{P}(\{x_i\},h)$,
the result does not depend on $t_s$, so that we set $t_s=0$, yielding
\be \label{response-rel}
\chi_p(t) = \left\langle B_p(t)\,
\frac{\partial \ln\mathcal{P}}{\partial h}(\{x_i(0)\},0)\right\rangle\;.
\ee

\subsection{Properties of the phase-space distribution}

\subsubsection{Uniform distribution on energy shells}

In order to go beyond the formal expression (\ref{response-rel-ttw})
of the response function, we need to choose a specific form of the
distribution $\mathcal{P}(\{x_i\},h,t_s)$.
We first consider the case when the distribution
only depends on the total energy $E_h$, namely
$\mathcal{P}(\{x_i\},h,t_s) = Z(t_s)^{-1} \exp[-\theta(E_h,t_s)]$, with
$Z(t_s)$ being the normalization constant.
The distribution $\mathcal{P}(\{x_i\},h,t_s)$ is thus uniform over constant
energy surfaces in phase-space for all times.
A linear time-independent $\theta(E,t_s)=\beta E+\theta_0$
corresponds to the equilibrium canonical ensemble.
However, we consider here the more general case of a regular function
$\theta(E,t_s)$ monotonically increasing with the total energy.
It is easy to check that for $h=0$, $\partial \ln Z/\partial h
= \langle M \theta'\rangle=0$,
so that $\partial \ln\mathcal{P}/\partial h=M \theta'$,
where $\theta'$ is the derivative of $\theta$ with respect to the total energy.
For a macroscopic system, the average in Eq.~(\ref{response-rel-ttw})
is dominated by the most probable energy level $E^*(t_s)$.
From a saddle-point evaluation, we obtain
\be
\chi_p(t,t_s) = \frac{\partial \theta}{\partial E}(E^*(t_s),t_s)\, C_p(t,t_s).
\ee
Hence a fluctuation-dissipation temperature
\be
T_\mathrm{FD}(t_s) = \left(\frac{\partial \theta}{\partial E}(E^*(t_s),t_s)\right)^{-1},
\ee
independent of the observable, can be defined.
As the value $E^*(t_s)$ maximizes the energy distribution $\rho(E,t_s) \propto
\exp[S(E)-\theta(E,t_s)]$, where $S(E)$ is the microcanonical entropy,
it turns out that $\partial \theta/\partial E(E^*(t_s),t_s)=S'(E^*(t_s))$,
so that the standard definition of temperature is recovered.

\subsubsection{Beyond uniformity: the lack of entropy}

In a more general situation, the distribution $\mathcal{P}(\{x_i\},h,t_s)$
is not uniform over the shells of constant energy, and the above simplification
does not occur, leading generically to an observable dependence of the
fluctuation-dissipation temperature
\footnote{If the FDR is non linear, the fluctuation-dissipation temperature
is not even defined for a given observable.}.
The above remarks suggest that this dependence on the observable
could be related to a macroscopic quantity,
namely the Shannon entropy difference between the stationary state and the
equilibrium state with the same average energy. When the distribution is
uniform over the most probable energy shell, the entropy is maximal, 
so that a non-uniform state necessarily corresponds
to a lower entropy. The entropy difference may thus be interpreted as
a measure of the deviation from equilibrium.

In the rest of this section, we focus on steady-state distributions
in order to simplify the notations, but our argument can
straightforwardly be extended to situations where the distribution
$\mathcal{P}(\{x_i\},h,t_s)$ depends on time.
As a general framework, we consider in the following a class of stochastic
markovian models, where an energy $E=\sum_{i=1}^N \ve_h(x_i)$ is
exchanged in a random way between the internal
degrees of freedom. Either the internal dynamics, or in more realistic
scenarios additional external sinks and sources, drive the system into
a nonequilibrium steady state. The resulting drive can be encompassed
by a dimensionless parameter $\gamma$,
like a normalized temperature difference or external force.
Note that in some cases the parameter
$\gamma$ may be the square of the physical driving
force, as we define $\gamma$ as the order of magnitude of the
leading correction to the equilibrium distribution (see below).
In the absence of driving ($\gamma=0$), detailed balance is satisfied
and the system is described by an equilibrium distribution
\begin{equation}
\mathcal{P}_\mathrm{eq}(\{x_i\},h)=Z_N^{-1}\exp\left(-\beta \sum_{i=1}^N \ve_h(x_i)\right)
\end{equation}
where $\beta=1/T$ is the inverse temperature of the thermal bath,
and $Z_N$ is the canonical partition function.

As a simplification, we assume that the $N$-body steady-state distribution
$\mathcal{P}(\{x_i\},h)$ can be factorized according to
$\mathcal{P}(\{x_i\},h) = \prod_{i=1}^N p(x_i,h)$, meaning that
the degrees of freedom are statistically independent.
The system can thus be fully described by means of the single-variable
probability distribution $p(x,h)$.
We now consider the small driving limit $|\gamma| \ll 1$, and expand
$p(x,h)$ around the equilibrium distribution
$p_\mathrm{eq}(x,h) = Z_1^{-1}\exp[-\beta \ve_h(x)]$ as
\begin{equation} \label{dist-dev}
p(x,h) = p_\mathrm{eq}(x,h) \left[1+\gamma F(\ve_h(x))+\mathcal{O}(\gamma^2)\right]\;.
\end{equation}
Such a perturbation is consistent with some recent generic results
on nonequilibrium distributions \cite{Komatsu}.
The constraints of normalization of $p(x,h)$ and $p_\mathrm{eq}(x,h)$
imply that $\langle F(\ve)\rangle_\mathrm{eq}=0$,
where $\ve$ is a shorthand notation for $\ve_h(x)$ and
$\langle \cdots \rangle_\mathrm{eq}$ denotes the equilibrium average.
If $p(x,h)$ follows Eq.~(\ref{dist-dev}), the factorized
$N$-body distribution $\mathcal{P}(\{x_i\},h)$ is generically not
a function of the total energy $E=\sum_{i=1}^N \ve_h(x_i)$, so that
$\mathcal{P}(\{x_i\},h)$ is not uniform over the shells of constant energy.
It follows that the nonequilibrium Shannon entropy is
lower than the entropy of the reference equilibrium state
having the same energy. 
Accordingly, the entropy difference between the equilibrium and nonequilibrium
states with the same average energy provides a
characterization of the deviation from equilibrium.
To determine the entropy difference,
we compute the average energy $E(\beta,\gamma)$ of the
out-of-equilibrium system, and we evaluate the temperature $\beta^*$ for
which $E(\beta,\gamma)=E_\mathrm{eq}(\beta^*)$, where
$E_\mathrm{eq}(\beta^*)$ is the equilibrium energy at temperature $\beta^*$.
As the distribution $\mathcal{P}(\{x_i\},h)$ is factorized,
the Shannon entropy of the whole system is simply computed
as the sum of the entropies associated to each variables
$x_i$. Hence only the Shannon entropy associated to a single degree of freedom,
\be
S=-\int dx\, p(x,h)\ln p(x,h)\;,
\ee
needs to be computed.
We denote by $S_\mathrm{eq}(\beta)$ the equilibrium entropy at temperature
$\beta$, and by $S(\beta,\gamma)$ the entropy in the nonequilibrium
steady-state characterized by $\beta$ and $\gamma$.
We then define the entropy difference $\Delta S$ per degree of freedom through
the relation
\be
\Delta S = S_\mathrm{eq}(\beta^*) - S(\beta,\gamma)\;.
\ee
A rather straightforward calculation yields (see Appendix \ref{app-deltaS}
for details):
\be\label{ds}
\Delta S = \frac{\gamma^2}{2} \left( \left< F\left(\ve\right)^2 \right>_\mathrm{eq}
- \frac{\langle \ve F\left(\ve\right) \rangle_\mathrm{eq}^2}{\langle \ve^2 \rangle_\mathrm{eq} - \langle \ve \rangle_\mathrm{eq}^2} \right)\;.
\ee
We have checked that $\Delta S\ge 0$ as expected
(see Appendix \ref{app-deltaS}),
even though this property does not appear explicitly
in Eq.~(\ref{ds}).
In the case of a linear $F(\ve)$, one finds $\Delta S = 0$,
which can be understood from the fact that the distribution $p(x,h)$
can be recast into an effective equilibrium form --see Eq.~(\ref{dist-dev}).
Considering now a generic function $F(\ve)$, we parameterize it as
\be
F(\ve) = a+b\ve+\eta f(\ve)\;,
\ee
where $\eta$ characterizes the amplitude of the nonlinearity.
The normalization condition $\langle F(\ve)\rangle_\mathrm{eq}=0$
fixes the value of the parameter $a$. We then obtain the generic result
\be \label{ds-perturb}
\Delta S=\gamma^2 \eta^2 \omega\;,
\ee
where $\omega$ is a constant which depends on the detailed expression
of the functions $f(\ve)$ and $\ve_h(x)$.
As an example, considering a nonlinearity of the form $f(\ve)=\ve^2$
and a zero-field local energy $\ve_0(x)=\frac{1}{2}x^2$, one finds
\be
\Delta S = \frac{3\gamma^2 \eta^2}{4\beta^4}\;.
\ee

\subsection{Fluctuation-dissipation relation and effective temperatures}

We now proceed to derive the FDR associated to the variable $B_p$,
and to analyze the dependence of the effective temperature on $p$.
To this aim, we further restrict the class of models considered
by making an additional simplification.
Namely, we assume that each time a variable $x_i$ is modified by a
dynamical event, its new value is decorrelated from the previous one.
Examples of models satisfying this assumption are given below
(see also \cite{Bertin-temp}).
Qualitatively, such an assumption can be interpreted as a coarse-graining
of the dynamics on a time scale of the order of the correlation time of the
system. Note that we focus here on systems
with a single relaxation time scale, so that the present approach does not
necessarily apply to systems with a more complex dynamics involving
different time scales, like glassy systems.
This assumption of local decorrelation implies that the correlation
function $C_p(t)$ is proportional
to the persistence probability $\Phi(t)$ \cite{Bertin-temp}, defined as
\begin{equation}
\Phi(t)=\left\langle\frac{1}{N}\sum_{i=1}^N \phi_i(t)\right\rangle,
\end{equation}
where the history-dependent local random variables $\phi_i(t)$ are equal
to $1$ if no redistribution involving site $i$ occurred between $t=0$ and
$t$, and are equal to $0$ otherwise. As a consequence we can express the
different averages involved in expressions (\ref{def-corr}) and
(\ref{response-rel}) in terms of $\Phi(t)$.

We expand the local energy $\ve_h(x)$ for small field $h$, namely
$\ve_h(x)=\ve_0(x)-h\psi(x) + \mathcal{O}(h^2)$,
assuming $\psi(x)$ to be an odd function.
It follows that the observable $M$, conjugated to the field $h$, is defined as
\be
M = \sum_{i=1}^N \psi(x_i) \;.
\ee
For the family of observables $B_p$, we choose the following definition:
\begin{equation}
B_p = \sum_{i=1}^N x_i^{2p+1}
\end{equation}
with $p\ge 0$ an integer number.
In this way, $B_p$ has a zero mean value in the absence of the field, which
slightly simplifies the calculations.

Using the fact that the random  variables $x_i$ and $x_j$ are independent
for $i\neq j$, and that $\langle B_p\rangle=\langle M\rangle=0$,
one can simplify the expression (\ref{def-corr}) for the correlation
function, leading to
\be
C_p(t) = N\langle x(t)^{2p+1} \psi(x(0)) \rangle \;.
\ee
Further using the assumption of decorrelation by the local dynamical events,
we get
\be
C_p(t) = N\langle x^{2p+1} \psi(x)\rangle\, \Phi(t)\;.
\label{cp-general}
\ee
The average in the last expression is performed on the steady-state
distribution.
From Eq.~(\ref{response-rel}), the response function can also be evaluated
using the decorrelation assumption, yielding
\be \label{chip-general}
\chi_p(t) = N[\beta \langle x^{2p+1}\psi(x)\rangle -
\gamma \langle x^{2p+1}\psi(x) F'(\ve_0) \rangle]\, \Phi(t)
\ee
where $\ve_0$ is a simplified notation for $\ve_0(x)$.
The average in the second term can be replaced by the equilibrium average,
neglecting higher order terms in $\gamma$.
One can then express $\Phi(t)$ as a function of $C_p(t)$ from
Eq.~(\ref{cp-general}), which leads to a FDR of the form (\ref{FDR-p}).
The corresponding temperature $\beta_p = T_p^{-1}$ is given by

\be\label{beta_p}
\beta_p = \beta-\gamma \frac{\langle x^{2p+1}\psi(x) F'(\ve_0)\rangle_\mathrm{eq}}{\langle x^{2p+1} \psi(x)\rangle_\mathrm{eq}}\;.
\ee
As anticipated, the $p$-dependence in Eq.~(\ref{beta_p}) does not cancel
in general, so that the temperature $T_p$ generically depends on the
observable.
A notable exception is the case of a linear $F(\ve)$, namely $F(\ve)=a+b\ve$,
where the effective temperature $\beta_p=\beta-\gamma b$ 
is observable independent.

When $F(\ve)$ has a nonlinear contribution, parameterized as
$F(\ve) = a+b\ve+\eta f(\ve)$, the temperature difference between
two distinct observables
is proportional to the amplitude $\eta$ of the nonlinearity.
More precisely, one finds from Eq.~(\ref{beta_p}) that
\be \label{betap-beta0}
\beta_p-\beta_0 = \gamma \eta \left[\frac{\langle x\psi(x) f'(\ve_0)\rangle_\mathrm{eq}}{\langle x \psi(x)\rangle_\mathrm{eq}}-
\frac{\langle x^{2p+1}\psi(x) f'(\ve_0)\rangle_\mathrm{eq}}{\langle x^{2p+1} \psi(x)\rangle_\mathrm{eq}}\right]\;.
\ee
On the other hand, we have seen in Eq.~(\ref{ds-perturb}) that the
lack of entropy $\Delta S$ is also a measure of the nonlinearity amplitude
$\eta$.
Hence it is natural to try to obtain a quantitative relation between
$\beta_p-\beta_0$ and $\Delta S$.
Starting from Eqs.~(\ref{ds-perturb}) and (\ref{betap-beta0}),
we get a relation of the form
\be \label{observable-entropy}
\frac{|\beta_p-\beta_0|}{\beta} = \kappa_p \sqrt{\Delta S}\;,
\ee
where we have introduced a dimensionless and positive constant $\kappa_p$.
This constant a priori depends on $p$, as well as on the functional forms
of $f(\ve)$ and of the local energy $\ve_h(x)$.
Note however that $\kappa_p$ does not depend on $\gamma$ and $\eta$.
In the following section, we give two examples of models for which
$\kappa_p$ can be determined exactly.

From the above analysis, it turns out that the dependence
of the fluctuation-dissipation temperature on the choice of observable
is a direct measure of the deviation from equilibrium.
As already mentioned, the above argument can be generalized to
time-dependent probability distributions.
In this case, $\beta_p$, $\kappa_p$ and
$\sqrt{\Delta S}$ may all depend on time.
Let us however emphasize again that the main assumptions made, namely that
the distribution remains close to some equilibrium state,
and that local decorrelation occurs in a single step, may not apply
to glassy systems.

\section{Illustration on simple models}
\label{sec-models}

\subsection{A simple energy transfer model on a ring}
\label{sec-ring}

\subsubsection{Model and steady-state solution}

\begin{figure}[t]
\centering\includegraphics[width=0.85\columnwidth,clip]{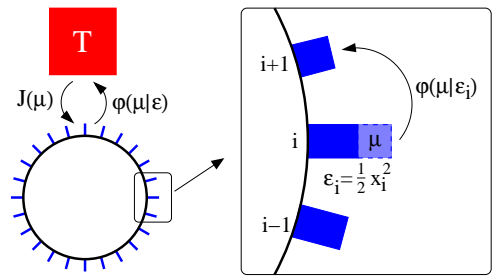}
\caption{Left: Scheme of the energy transport model on a ring in contact with a bath at temperature $T$. Energy is injected from the bath to the ring with rate $J(\mu)$ and dissipated from the ring to the bath with rate $\varphi(\mu|\ve)$. Right: Internal dynamics of the ring. An fraction $\mu$ of the local energy $\ve_i=x_i^2/2$ is transported from site $i$ to site $i+1$ on the ring according to the transport rate $\varphi(\mu|\ve_i)$.}
\label{fig-ring}
\end{figure}

To illustrate the above results let us consider as a first example an energy transfer model on a ring geometry that is connected to a bath with temperature $T$  (see Fig.~\ref{fig-ring}).
The model is defined on a one-dimensional lattice with periodic boundary conditions.
To every site $i$, $i=1\ldots N$, is attached a real quantity $x_i$,
associated to a local energy $\ve_i=\frac{1}{2}x_i^2$.
A fraction $\mu$ of the local energy $\ve_i$ is transferred
from site $i$ to site $i+1$, according to the site independent rate
\begin{equation}\label{trans-rate}
\varphi(\mu|\ve_i)=v(\mu)\frac{g(\ve_i-\mu)}{g(\ve_i)},
\qquad g(\rho)=\rho^{\delta-1}\;,
\end{equation}
with $\delta>0$, and $v(\mu)$ an arbitrary positive function.
After the transport, the new variables denoted as $x_i'$ and $x_{i+1}'$
take the values
\be
x_i'=\pm\sqrt{x_i^2-2\mu} \;, \qquad x_{i+1}'=\pm\sqrt{x_{i+1}^2+2\mu}\;,
\label{new-var}
\ee
with equiprobable and uncorrelated random signs. We consider a continuous time dynamics, where sites are updated in an asynchronous way.
As can be easily checked, these transport rules locally conserve the energy.
The choice of the function $g(\rho)$ entering the transport rates also ensures that the system remains homogeneous (no condensation occurs) \cite{Evans}.

In addition, each site $i$ of the system is also connected to an external
heat bath at temperature $T$, according to the following dynamics.
An amount of energy $\mu$ is injected from the bath with a probability rate
$J(\mu)$ given by
\be
J(\mu) = v(\mu) e^{-\mu/T}\;.
\ee
Energy is transferred back to the bath with the same energy transport rate $\varphi(\mu,\ve_i)$ as for the internal transport.

Given this dynamics, the steady-state probability distribution for a microscopic state $\{x_i\}$ takes the factorized form
\begin{equation}\label{dist-can-toy}
\mathcal{P}_0(\{x_i\})=\frac{1}{Z_N} \prod_{i=1}^N \left\{|x_i|
g\left(\frac{x_i^2}{2}\right)\right\} e^{-\frac{1}{T}\sum_i\frac{1}{2}x_i^2}\;,
\end{equation}
with $Z_N$ the normalization factor of the distribution, and where the index
$0$ indicates a zero-field dynamics.

In the following, we show that the temperature defined from the
FDR does not necessarily coincide with the bath temperature,
consistently with Sect.~\ref{sec-frame} and
with the results obtained in a similar model \cite{Bertin-temp}.
The two temperature definitions become equivalent
only for the special choice $\delta=1/2$ in the transport rates
(\ref{trans-rate}), when the probability distribution has an equilibrium form.

\subsubsection{Fluctuation-dissipation relations}
\label{sec-response}

To relate the spontaneous fluctuations present in this stochastic system to the response of an observable to a small perturbation, let us introduce an external field $h(t)$ perturbing the system. A natural way to couple the field to the system is to add to the energy a linear term proportional to the external field:
\begin{equation}\label{h-coupling}
E_h=\sum_{i=1}^N \frac{1}{2}x_i^2-h\sum_{i=1}^N x_i+\frac{Nh^2}{2}=
\sum_{i=1}^N \frac{1}{2}(x_i-h)^2\;,
\end{equation}
where we included for convenience an additional shift to the energy $Nh^2/2$
which is only changing the reference of the energy scale without
loss of generality.
Note that Eq.~(\ref{h-coupling}) implies $\psi(x)=x$ and $M=\sum_{i=1}^N x_i$.
Introducing the new variables $v_i=x_i-h$ we ask that they obey
the same dynamics as the former variables $x_i$, that are given in Eqs.~(\ref{trans-rate}) and (\ref{new-var}).
Further we assume the same protocol for the externally applied perturbation as described in the previous section: the field $h(t)$ is non-zero at times $t<0$, but small in comparison to the mean value $\langle x\rangle$ of the variables.
We assume that the nonequilibrium steady state is established for $t<0$. At time $t=0$ the field is switched off in order to analyze the response of an observable $B_p(t)$ to this variation of the field.

Following Sect.~\ref{sec-frame}, we consider the observables
$B_p$ defined as $B_p=\sum_{i=1}^N x_i^{2p+1}$,
with $p$ a positive integer number.
Given that the distribution (\ref{dist-can-toy}) is factorized,
the results of Sect.~\ref{sec-frame} can be applied, and
the steady-state correlation function $C_p(t)=\langle B_p(t)M(0) \rangle$
(we recall that $\langle B_p\rangle=\langle M\rangle=0$)
is given by
\begin{equation} \label{cp-ring}
C_p(t)= N \langle x(t)^{2p+1} x(0)\rangle\;.
\end{equation}
To obtain the general formulation of the response
function we use expression (\ref{response-rel})
with $\mathcal{P}(\{x_i(0)\},h)$ being the distribution
for the nonequilibrium steady state in the presence of the field $h$.
This distribution is given by (\ref{dist-can-toy}) with respect to the
new variables $v_i=x_i-h$, namely
$\mathcal{P}(\{x_i(0)\},h)=\mathcal{P}_0(\{v_i(0)\})$, meaning that
the dynamics of the variables $\{x_i\}$ in the presence of the field $h$
can be effectively described as a zero-field dynamics, once expressed
in terms of the variables $\{v_i\}$.

For arbitrary values of the integer $p\ge 0$, we obtain the following 
relation between the response and the correlation in the system for the
observable $B_p(t)$ (details are given in Appendix \ref{app-fdr}):
\begin{eqnarray} \label{chip-ring}
\chi_p(t)=\frac{2p+1}{2(p+\delta)}\frac{1}{T} \,C_p(t)\;.
\end{eqnarray}
The temperature defined by the fluctuation-dissipation relation generically
depends on $p$ and therefore on the observable chosen
\begin{equation}\label{temp-fd}
T_p=\frac{2(p+\delta)}{(2p+1)} \, T\;.
\end{equation}
Only for $\delta=1/2$, when the energy distribution (\ref{dist-can-toy})
is uniform, the temperature takes independently of the observable the value
$T_p=T$.
But for non-uniform energy distributions, the temperature determined
from the slope of the FDR depends on the observable and is therefore
not well-defined.

To study the weakly nonequilibrium regime, we consider values of $\delta$
close to the equilibrium value $\delta=1/2$, namely
$\delta=1/2+\gamma$ with $|\gamma|\ll 1$. 
We find for the linear correction $F(\ve)$ to the probability distribution
$p(x,h)$, as defined in Eq.(\ref{dist-dev}), the following expression:
\begin{eqnarray}
F(\varepsilon)&=&\ln \varepsilon + C_\beta \;,
\end{eqnarray}
where $C_\beta=\ln\beta-\psi_0(\frac{1}{2})$, and
$\psi_0$ denotes the digamma function, defined as the logarithmic derivative of the Euler gamma function (numerically, $\psi_0(\frac{1}{2})\approx-1.963$).
Knowing $F(\varepsilon)$ then allows for the determination of all important
quantities, necessary for the establishment of the crucial relation
(\ref{observable-entropy}).
Thus we end up with a quantitative expression of the observable-dependence
of the FDR-temperature in terms of the lack of entropy. 
From Eq.~(\ref{beta_p}), the observable dependence can be expressed through
\begin{equation}
\frac{|\beta_p-\beta_0|}{\beta}=|\gamma| \frac{4p}{2p+1}\;.
\end{equation}
Besides, the entropy difference can be evaluated from Eq.~(\ref{ds}), yielding
\begin{eqnarray}
\Delta S&=&\frac{\gamma^2}{2}\left[\langle(\ln\ve+C_\beta)^2\rangle-2\beta^2\langle\ve(\ln\ve+C_\beta)\rangle^2\right]\nonumber\\
&=&\frac{\gamma^2}{2}\left[\psi_0'\left(\frac{1}{2}\right)-2\right]\;,
\end{eqnarray}
where $\psi_0'$ is the derivative of $\psi_0$.
Numerically, we find $\Delta S/\gamma^2 \approx 1.467$.
From the relation (\ref{observable-entropy}), we then get
\begin{eqnarray}
\kappa_p&=&\frac{4p}{2p+1} \left[\frac{2}{\psi_0'\left(\frac{1}{2}\right)-2}\right]^{\frac{1}{2}}\;,
\end{eqnarray}
which is, as expected, independent of the physical parameters of the system,
like the driving $\gamma$ and the temperature $\beta$.

\subsubsection{Discussion of the transport model results}
\label{fdr1-discussion}

We learn from the study of this exactly solvable transport model several
interesting facts. First, this model illustrates explicitly how a
probability distribution that is not uniform on energy shells
(namely, $\delta\neq 1/2$) leads to a FDR-temperature which depends on the
observable.
The results for expression (\ref{observable-entropy}), that uniquely depend on the form of the probability distribution, do not depend on the energy flux in the system.
Therefore the energy flux in the bulk of the system does not play a crucial role for the results. We chose the rules of the dynamics such that the transport of energy is totally biased. But it is known that the symmetric case, where energy is transported with the same probability to the left or to the right, leads to exactly the same probability distribution for the microstates \cite{Evans}.
More precisely the distribution only depends on the transport rates, but the direction of the transport, that defines the total flux, is not of any influence.  

Further, it is interesting to compare the characterization in terms of
$\Delta S$ with that in terms of entropy production.
On general grounds, the entropy production $\sigma_s$ can be defined
from a balance equation involving rate of entropy change,
and the entropy fluxes with the reservoirs to which the system is connected:
\be\label{entropy-production}
\frac{dS}{dt} = \sum_{i=1}^n\frac{\mathcal{J}_i}{T_i}+\sigma_s\;,
\ee
where $\mathcal{J}_i$ is the energy flux exchanged with the $i^{\rm th}$
bath at temperature $T_i$.
In the present model, the steady-state implies $dS/dt=0$.
The system is in contact with a single bath, and the energy flux
$\mathcal{J}$ is zero. Thus the entropy production is also equal to zero.
This means that in the framework of this model the entropy production
cannot give any information about the observable dependence of the
effective temperature in the system. 

In the next section we present a different situation, where the nonequilibrium
character no longer results from an artificial bulk dynamics, but rather
from an external drive.

\begin{figure}[t]
\centering\includegraphics[width=6cm,clip]{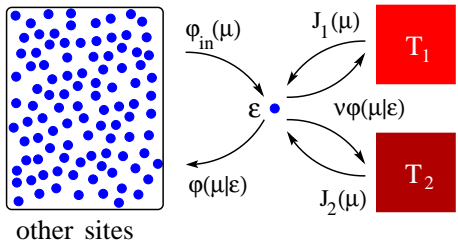}
\caption{Scheme of the fully connected model considered in
Sect.~\ref{sec-mean}. A single site contains an amount of energy $\varepsilon$.
It is in contact with two baths at different temperatures $T_1=\beta_1^{-1}$
and $T_2=\beta_2^{-1}$ and with the other sites.}
\label{fig-mf-model}
\end{figure}

\subsection{Fully connected model driven by two heat baths}
\label{sec-mean}

\subsubsection{Model and evolution equation}

In general we would expect that the non-uniformity of the
probability distribution results from the fact that the system
is externally driven into a nonequilibrium steady state,
for example by two reservoirs at different temperatures \cite{fdrprl}.
In the following we consider a model with $N$ fully connected sites,
associated to variables $x_i$,
such that the local energy $\ve_i=\frac{1}{2}(x_i-h)^2$
can be transferred between any pair of sites
and with two different thermal baths.
A sketch of the model is shown on Fig.~\ref{fig-mf-model}.
Energy transfers correspond to the dynamical rules (\ref{new-var})
in terms of variables $x_i$.
An amount of energy $\mu$ is transferred from an arbitrary site $i$,
with energy $\ve_i$, to any other site $j$ with a rate
\be \label{phi-mueps}
\varphi(\mu|\varepsilon_i) = \frac{g(\varepsilon_i-\mu)}{g(\varepsilon_i)}\;,
\qquad g(\rho) = \rho^{-\frac{1}{2}}\;.
\ee
Such a rate is similar to the rate (\ref{trans-rate}) for $\delta=\frac{1}{2}$
and $v(\mu)=1$.
The value $\delta=\frac{1}{2}$ is chosen such that equilibrium is recovered
when the two baths have the same temperature.
Energy is transferred to the baths with the same rate (\ref{phi-mueps})
as in the bulk, but weighted with an factor $\nu$ characterizing the coupling strength between
the baths and the system. The injection from the bath is defined
as the transfer, with a rate $\nu \varphi(\mu|\varepsilon)$,
from an equilibrated site having a distribution
$P_\mathrm{eq}(\varepsilon,\beta_{\alpha})$
at inverse temperature $\beta_{\alpha}$, leading to
\be
J_\alpha(\mu) = \nu\int_\mu^\infty d\varepsilon\, \varphi(\mu|\varepsilon)\,
P_\mathrm{eq}(\varepsilon,\beta_{\alpha}).
\ee
A straightforward calculation then yields

\be \label{injection}
J_\alpha(\mu)=\nu  e^{-\beta_\alpha \mu} \;.
\ee
At this stage, it is convenient to describe the dynamics in terms of the
local energy $\ve_i$ rather than with the variables $x_i$.
In the thermodynamic limit $N \to \infty$, the master equation
governing the $N$-body distribution can be recast into a nonlinear
evolution equation for the one-site probability distribution $P(\varepsilon)$,
namely
\begin{eqnarray}\label{P-evolution}
\frac{\partial P(\varepsilon,t)}{\partial t} &=& \int_0^\varepsilon d\mu\; (J_1(\mu)+J_2(\mu))P(\varepsilon-\mu,t)\nonumber\\
&&-\int_0^\infty d\mu\; (J_1(\mu)+J_2(\mu))P(\varepsilon,t)\nonumber\\
&&+(2\nu+1) \int_0^\infty d\mu\; \varphi(\mu|\varepsilon+\mu)P(\varepsilon+\mu,t)\nonumber\\
&&-(2\nu+1) \int_0^\varepsilon d\mu\; \varphi(\mu|\varepsilon)P(\varepsilon,t)\nonumber\\
&&+ \int_0^\varepsilon d\mu\; \varphi_\mathrm{in}(\mu,t) P(\varepsilon-\mu,t)\nonumber\\
&&- \int_0^\infty d\mu\; \varphi_\mathrm{in}(\mu,t) P(\varepsilon,t)\;.
\end{eqnarray}
The distribution $\varphi_\mathrm{in}(\mu,t)$ accounts for the energy transfer
coming from all the other sites, given by the averaged
transport rate $\varphi(\mu|\varepsilon)$:
\begin{equation} \label{phi-in-mu}
\varphi_\mathrm{in}(\mu,t)=\int_\mu^\infty d\varepsilon\; \varphi(\mu|\varepsilon)P(\varepsilon,t)\;.
\end{equation}
The local energy distribution $P(\ve,t)$ is related to the distribution
$p(x,h,t)$ through
\begin{equation}
p(x,h,t)=\frac{1}{2}P(\ve,t) \left|\frac{d\ve_h}{d x}\right|
\end{equation}
where $\ve=\ve_h(x)=\frac{1}{2}(x-h)^2$.

\begin{figure}[t]
\centering\includegraphics[width=8cm,clip]{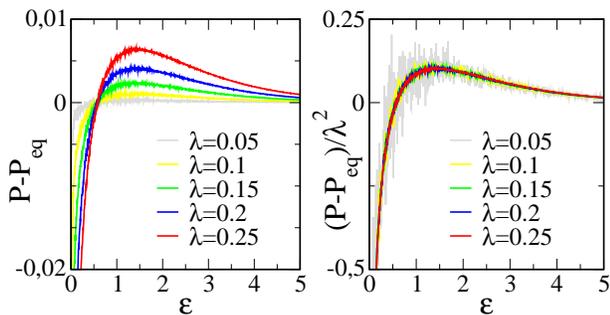}
\caption{Test of the analyticity of $P(\varepsilon)$. Numerical simulations
of the fully connected model show that the curves for
$[P(\varepsilon)-P_{\mathrm{eq}}(\varepsilon)]/\lambda^2$ match to a very good
accuracy for small values of $\lambda$, justifying the assumption of
analyticity of the $\lambda$-expansion made in Eq.~(\ref{dist-dev2}).
Parameter values: $\nu=1$, $\beta=1$, $T_\mathrm{max}=10^8$.}
\label{fig-lambda-comparison}
\end{figure}

\subsubsection{Stationary distribution for a small bath temperature difference}

\begin{figure}[t]
\centering\includegraphics[width=6.5cm,clip]{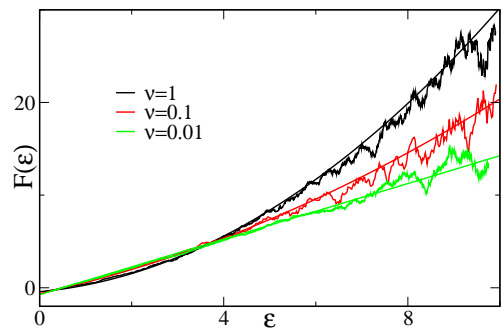}
\caption{The function $F(\ve)$ obtained by simulations (noisy lines) in comparison with the analytically obtained results for $F^{(2)}(\ve)$ (solid lines) for different values of $\nu$ ($\lambda=0.2$, $\beta=1$, $T_\mathrm{max}=10^7$).}
\label{fig-F-of-eps}
\end{figure}

In order to determine the steady-state distribution of the model,
we consider the case of a small temperature difference
$|\beta_1-\beta_2| \ll (\beta_1+\beta_2)/2$, and parameterize the bath
temperatures as $\beta_1=\beta (1-\lambda)$ and $\beta_2=\beta (1+\lambda)$,
with $\lambda \ll 1$.
We then assume that the stationary distribution $P(\varepsilon)$ has an
analytical expansion as a function of $\lambda$.
The linear term in $\lambda$ is excluded by a simple symmetry
argument: exchanging the two bath temperatures changes $\lambda$ into
$-\lambda$, but should not modify the distribution $P(\varepsilon)$
since the two heat baths play a symmetric role.
The leading correction should thus behave as $\lambda^2$, so that
$P(\varepsilon)$ can be written as
\begin{equation} \label{dist-dev2}
P(\varepsilon)=P_\mathrm{eq}(\varepsilon) \left[1+\lambda^2 F(\varepsilon)+\mathcal{O}(\lambda^4)\right]\;,
\end{equation}
in analogy to expression (\ref{dist-dev}) with $\gamma=\lambda^2$.
The distribution
\begin{equation}
P_\mathrm{eq}(\varepsilon)=\sqrt{\frac{\beta}{\pi}} \varepsilon^{-\frac{1}{2}} e^{-\beta\varepsilon}
\end{equation}
is the known equilibrium distribution, namely the stationary solution
of the equation (\ref{P-evolution}) for $\beta_1=\beta_2=\beta$.
The scaling form (\ref{dist-dev2}) is validated by numerical simulations, as
shown on Fig.~\ref{fig-lambda-comparison}.
The function $F(\varepsilon)$ determined from numerical data
is also shown in Fig.~\ref{fig-F-of-eps}
for different values of the coupling strength $\nu$.
The numerical results were obtained by directly simulating the stochastic
dynamics, on a system of size $N=102$. Such a relatively small size
allows for long time averaging, over time $T_\mathrm{max}$ of the order of
$10^7$ or $10^8$, in order to reach a sufficient statistics.
Note that the time unit is defined in such a way that all sites have
typically experienced about one redistribution event in a unit of time.

It can be checked that Eq.~(\ref{P-evolution}) has no exact solution
involving a finite polynomial function $F(\ve)$. To find the best polynomial approximation $F^{(L)}(\ve)$ at a given order $L$ we use a variational 
procedure, as detailed in Appendix~\ref{app-F}.
We approximate the function $F(\ve)$ by a polynomial of order $L$,
$F^{(L)}(\ve)=\sum_{n=0}^L a_n^{(L)} \beta^n\ve^n$.
The best approximation is then obtained analytically by minimizing the error,
under the constraints of normalization and zero net flux with the baths,
in the evolution equation (\ref{P-evolution}) linearized with respect
to the parameter $\gamma=\lambda^2$.
The error is defined as the equilibrium average of the square
of the r.h.s.~in the linearized equation.

If we use for example $F^{(2)}(\ve)$ to approximate $F(\ve)$, the normalization constraint and the constraint of zero net flux in the system yields for the coefficients $a_0^{(2)}$ and $a_1^{(2)}$ (see Appendix~\ref{app-F}):
\begin{equation}\label{a0a1}
a_1^{(2)}=\frac{3}{2}-4a_2^{(2)} \;, \quad
a_0^{(2)}=-\frac{3}{4}+\frac{5}{4}a_2^{(2)}\;.
\end{equation}
Hence the only remaining free parameter is $a_2^{(2)}$.
Minimizing the error with respect to $a_2^{(2)}$ yields an analytic
expression for the coefficient $a_2^{(2)}$ 
as a function of the coupling strength $\nu$ (see Appendix \ref{app-F})
\begin{equation}\label{a2}
a_2^{(2)}(\nu)=\frac{3\nu(7+37\nu)}{13+136\nu+358\nu^2}\;.
\end{equation}
Note that in the limit of small $\nu$ this expression vanishes linearly in $\nu$, whereas the coefficients $a_0^{(2)}$ and $a_1^{(2)}$ take the finite values
$-\frac{3}{4}$ and $\frac{3}{2}$ respectively.

\begin{figure}[t]
\centering\includegraphics[width=6.5cm,clip]{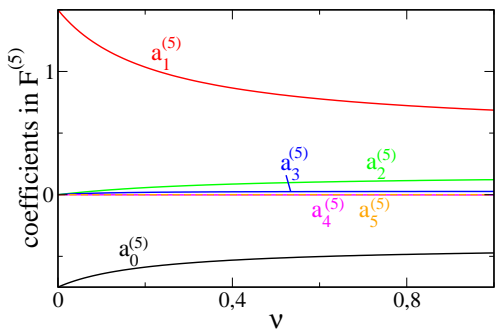}
\caption{$\nu$-dependence of the coefficients in $F^{(5)}(\ve)$ (see text).}
\label{fig-coeff}
\end{figure}

\begin{figure}[t]
\centering\includegraphics[width=6.5cm,clip]{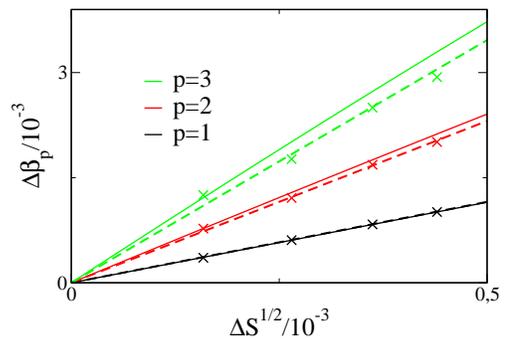}
\caption{Parametric plot in $\nu$ of $\Delta \beta_p=|\beta_p-\beta_0|/\beta$ versus $\sqrt{\Delta S}$ obtained using $F^{(5)}(\ve)$, either fitted to numerical data ($\times$) or calculated in the analytical approximation (solid lines).
Dashed lines: Eq.~(\ref{observable-entropy}) with $\kappa_p=4p/\sqrt{3}$.
Simulation parameters: $\lambda=0.05$, $T_\mathrm{max}=10^7$.}
\label{fig-dS}
\end{figure}

Similarly, one can find analytically higher order approximations for $F(\ve)$.
Through this procedure we find that for approximations with $L>2$ the coefficients $a_k^{(L)}$, $k>2$, in the expansion
are numerically small, as illustrated in Fig.~\ref{fig-coeff}(c).
A second order polynomial is thus already a good approximation of $F(\ve)$ for $\nu\lesssim 1$ [see Fig.~\ref{fig-F-of-eps}(b)].
Taking into account higher order terms in $F(\ve)$, we find that
the relation (\ref{observable-entropy}) between the observable
dependence and the entropy difference is satisfied to a good accuracy,
as seen in Fig.~\ref{fig-dS}.
We have also checked for $L\leq 5$ that in the limit $\nu \to 0$,
the coefficients $a_k^{(L)}$, $k>1$ vanish while $a_0^{(L)} \to -\frac{3}{4}$
and $a_1^{(L)} \to \frac{3}{2}$. 

\subsubsection{Observable-dependence and its relation to the lack of entropy}

In terms of the variable $x$, the stationary one-body distribution $p(x,h)$
reads, to second order in $\lambda$
\begin{equation}
p(x,h)=\sqrt{\frac{\beta}{2\pi}}e^{-\frac{1}{2}\beta(x-h)^2}
\left[1+\lambda^2 F\left(\frac{1}{2}(x-h)^2\right)\right]\;.
\end{equation}
Following our standard procedure, we introduce the response $\chi_p(t)$ of the
observable $B_p$ to a small perturbing field $h$.
From Sect.~\ref{sec-frame} we know how to obtain the FDR-temperature and the entropy difference using the correction $F(\ve)$. As shown above $F(\ve)$ can be well approximated by a quadratic function $F^{(2)}(\ve)=a_0^{(2)}+a_1^{(2)}\beta\ve+a_2^{(2)}\beta^2\ve^2$. Consequently we obtain for the observable-dependent inverse temperature, using Eq.~(\ref{beta_p}) with $\psi(x)=x$ 
\begin{eqnarray}
\beta_p&=&\beta-\lambda^2\frac{\langle\ve^{p+1}(a_1^{(2)}\beta+2 a_2^{(2)}\beta^2\ve)\rangle_\mathrm{eq}}{\langle\ve^{p+1}\rangle_\mathrm{eq}}
\end{eqnarray}
leading to
\begin{eqnarray}
\frac{|\beta_p-\beta_0|}{\beta}&=& 2 p \lambda^2 |a_2^{(2)}| \;.
\end{eqnarray}
The calculation of $\Delta S$ using Eq.~(\ref{ds}) with $F^{(2)}(\ve)$ is
also simple, involving averages of different powers of the energy,
and the result takes the form
\begin{equation}
\Delta S=\frac{3}{4}\lambda^4 (a_2^{(2)})^2 \;.
\end{equation}
Thus we obtain for the relation between the
entropy difference $\Delta S$ and observable-dependence of the temperature
\begin{equation}\label{obs-S-mean}
\frac{|\beta_p-\beta_0|}{\beta}=\frac{4p}{\sqrt{3}} \sqrt{\Delta S}
\end{equation}
in agreement with the general results presented in Section \ref{sec-frame}
--see Eq.~(\ref{observable-entropy}).

\subsubsection{Discussion of the fully-connected model}

The study of this model shows that observable dependence occurs as soon as the dynamics results in non-uniform probability distributions for the microstates.  
Further it is again possible to use the proposed general approach to characterize this non-uniformity by the entropy difference $\Delta S$.
For small driving we find a direct relation between this entropy difference and the observable dependence (\ref{obs-S-mean}).
The proportionality factor turns out to be independent of the coupling strength and the driving parameter. 
However, in the zero coupling limit, where $F(\ve)$ becomes linear, both $\Delta S$ and $|\beta_p-\beta_0|/\beta$ vanish, meaning that for small coupling we expect no observable dependence.

To compare these results with the information given by the entropy production
$\sigma_s$ we calculate the total energy fluxes caused by the contact
to the different baths.
We denote as $\mathcal{J}_\mathrm{out}$ the total energy flux transferred
from the systems to both heat baths, and by $\mathcal{J}_\mathrm{in}$
the total flux injected by the baths. In steady state, one has
$|\mathcal{J}_\mathrm{out}|=|\mathcal{J}_\mathrm{in}|$.
The flux $\mathcal{J}_\mathrm{in}$ is computed as
\be
\mathcal{J}_\mathrm{in} = \int_0^\infty d\mu (J_1(\mu)+J_2(\mu))\mu \;.
\ee
Expanding the above integral to second order in $\lambda$, we obtain
\be
\mathcal{J}_\mathrm{in} = \frac{\nu}{\beta^2}\left[2+6\lambda^2+\mathcal{O}(\lambda^4)\right]\;.
\ee
A similar calculation yields for the net energy fluxes $\mathcal{J}_{\alpha}$
exchanged by the bath $\alpha$ with the system
\begin{eqnarray}
\mathcal{J}_1&=&\int_0^\infty d\mu J_1(\mu)\mu-\frac{1}{2}|\mathcal{J}_\mathrm{out}|\nonumber\\
&=& 2\frac{\lambda\nu}{\beta^2}+\mathcal{O}(\lambda^3)\\
\mathcal{J}_2&=& \int_0^\infty d\mu J_2(\mu)\mu-\frac{1}{2}|\mathcal{J}_\mathrm{out}|\nonumber\\
&=& -2\frac{\lambda\nu}{\beta^2}+\mathcal{O}(\lambda^3)\;.
\end{eqnarray} 
These results lead to an entropy production
\begin{eqnarray}
\sigma_s=-(\beta_1 \mathcal{J}_1+\beta_2 \mathcal{J}_2)
= \frac{4\nu \lambda^2}{\beta}\;.
\end{eqnarray}
Hence we can relate $|\beta_p-\beta_0|$ to the entropy production as follows:
\begin{equation} \label{bpb0-sigma}
 \frac{|\beta_p-\beta_0|}{\beta}=\zeta(\nu)\frac{p \beta}{2} \sigma_s
\end{equation}
with $\zeta(\nu)= |a_2^{(2)}(\nu)|/\nu$, the coefficient $a_2^{(2)}(\nu)$
being given in Eq.~(\ref{a2}).
Consequently in the framework of this model it is possible to relate the dependence of the temperature on the observable to the entropy production.

The quantity $|\beta_p-\beta_0|/\beta$ results linear in $\sigma_s$
in contrast with the characterization through the entropy difference
--see Eq.~(\ref{obs-S-mean}).
Similarly to Eq.~(\ref{obs-S-mean}) the proportionality factor
does not depend on the bath temperature difference $\lambda$.
But contrary to the characterization via the lack of entropy,
the coupling strength $\nu$ now enters into relation (\ref{bpb0-sigma}).
Hence the entropy difference $\Delta S$ seems to be more directly related
to $|\beta_p-\beta_0|$ than the entropy production $\sigma_s$.
Note however that in the small coupling limit $\zeta(\nu)$ becomes a constant,
so that the dependence upon the coupling constant disappears in this limit.

\subsection{Slow relaxation model} 
\label{sec-slow}

\subsubsection{Model and time dependent probability distribution}

\begin{figure}[t]
\centering\includegraphics[width=7cm,clip]{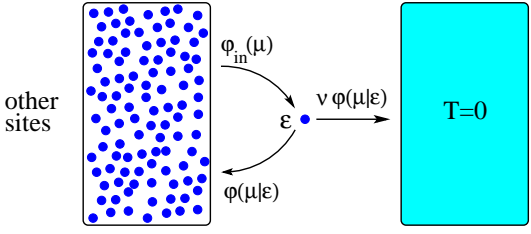}
\caption{Scheme of the fully-connected model in contact with a bath at zero temperature.}
\label{fig-zero}
\end{figure}

The former results seem to indicate that in the case of driven systems the observable dependence is a direct result of the non-uniformity of the phase space distribution.
In Sect.~\ref{sec-frame}, we argued that such results also hold in
the time-dependent case.
In the following we will investigate a similar fully connected model as in the above example, but put into contact with a single heat bath at zero temperature (see Fig.~\ref{fig-zero}).
Interestingly this model can be solved exactly in the non-stationary regime.
In this case the evolution equation for the probability distribution of the microstates in the thermodynamic limit reads
\begin{eqnarray}\label{P-evolution-slow}
\frac{\partial P(\varepsilon,t)}{\partial t}&=&(\nu+1) \int_0^\infty d\mu\; \varphi(\mu|\varepsilon+\mu)P(\varepsilon+\mu,t)\nonumber\\
&&-(\nu+1) \int_0^\varepsilon d\mu\; \varphi(\mu|\varepsilon)P(\varepsilon,t)\nonumber\\
&&+ \int_0^\varepsilon d\mu\; \varphi_\mathrm{in}(\mu,t) P(\varepsilon-\mu,t)\nonumber\\
&&- \int_0^\infty d\mu\; \varphi_\mathrm{in}(\mu,t) P(\varepsilon,t)\;,
\end{eqnarray}
with $\varphi_\mathrm{in}(\mu,t)$ given in Eq.~(\ref{phi-in-mu}).
Using as an ansatz the time-dependent Gibbs distribution
$P(\ve,t)=\sqrt{\beta(t)/\pi\ve}\, e^{-\beta(t) \ve}$
in Eq.~(\ref{P-evolution-slow}), we find the following
differential equation, which should be valid for all $\ve>0$:
\begin{equation}
\frac{1}{2}\frac{\dot{\beta}}{\beta}-\dot{\beta}\ve=\frac{\nu}{\beta}
-2\nu\ve\;,
\end{equation}
where $\dot{\beta}$ denotes the time derivative of $\beta(t)$.
One can easily check that the above equation implies $\dot{\beta}=2\nu$.
Hence, starting at $t=0$ from an equilibrium distribution at temperature
$T(0)=\beta_\mathrm{init}^{-1}$,
the probability distribution is for $t>0$
a Gibbs-like distribution at temperature $T(t)=\beta(t)^{-1}$ given by
\be
T(t)= \frac{1}{\beta_\mathrm{init} + 2\nu t} \;.
\ee
Once expressed in terms of the variable $x$, the distribution reads
\be \label{prel}
p(x,h,t)=\frac{1}{\sqrt{2\pi T(t)}} \, \exp\left[-\frac{(x-h)^2}{2T(t)} \right] \;.
\ee
The entropy difference $\Delta S$ is thus equal to zero for all times.

\subsubsection{Fluctuation dissipation relation}

From Eq.~(\ref{response-rel-ttw}),
and taking into account the factorization property,
the response $\chi_p(t,t_s)$ of the observable $B_p$ is given for $t>t_s$ by
\be \label{response-t}
\chi_p(t,t_s) = N\left\langle x(t)^{2p+1}\,
\frac{\partial \ln p}{\partial h}(x(t_s),0,t_s)\right\rangle\;,
\ee
the average being computed at zero field. Therefore, we obtain the following result for the fluctuation-dissipation relation, using the probability density
$p(x,h,t_s)$ given by Eq.~(\ref{prel})
\begin{eqnarray}
\chi_p(t,t_s) &=& N\beta(t_s) \left\langle x(t)^{2p+1} x(t_s) \right\rangle \nonumber\\
&=&\beta(t_s) C(t,t_s)\;,
\end{eqnarray}
where $C(t,t_s)=N\langle x(t)^{2p+1} x(t_s) \rangle$ denotes the two time correlation function for the relaxation dynamics without field.
Thus the FDR defines an effective temperature that is independent of the observable. This result is consistent with the generic relation~(\ref{observable-entropy}) we obtained between the entropy difference and the observable dependence of the FDR (although $\kappa_p$ does not have here a well-defined value).

\subsubsection{Discussion of the relaxation model}

We could show within this model that even for relaxation dynamics,
the generic relation (\ref{observable-entropy}) that we derived for the
observable-dependence of the fluctuation-relation temperature still holds.
We find that the probability distribution has a Gibbs form for all times,
which results in zero entropy difference and no
observable-dependence of the fluctuation-dissipation temperature. 

Can we have similar predictions using the entropy production, as was the case
in the former model?  The definition of the entropy production
(\ref{entropy-production}) is not valid for a zero temperature bath.
Such a situation is however a theoretical idealization.
The entropy production can be evaluated for an arbitrarily small bath
temperature. In this limit the entropy production
becomes arbitrarily large, in contrast to the entropy difference which is
zero. Thus again, like in the example of the ring model, the entropy
production cannot be considered as a relevant characterization of the
observable-dependence of the fluctuation-dissipation temperature.

\section{Discussion and conclusion}
\label{sec-discuss}

In this work we were aiming for some general statements regarding the issue of observable dependent temperatures defined from fluctuation-dissipation
relations.
We studied two stochastic models with non-uniform probability distributions and another stochastic model in a relaxation regime, that exhibits a time-dependent distribution of Gibbs form.
These studies, complemented by the more general arguments developed
in Sect.~\ref{sec-frame}, support the view that the observable-dependence
of fluctuation-dissipation temperatures in driven systems results
from the non-uniformity of the phase space distribution. 

In order to find a characterization of the observable dependence we related the non-uniformity of microstate distribution on a given energy shell to a
quantity we call ``lack of entropy'' or ``entropy difference'' $\Delta S$,
namely the Shannon entropy
difference of the non-equilibrium system with respect to the equilibrium state
with the same average energy.
We generically found that the difference between the temperatures associated
to two different observables is proportional to the square-root of
$\Delta S$. This relation has been confirmed in the three explicit examples
studied.
In contrast, another quantity deeply rooted in nonequilibrium theory,
namely the entropy production, does not seem to provide a systematic
characterization of the dependence of the effective temperature upon
the observable. A summary of the results is presented in Table~\ref{table}.

\begin{table}[t]
\begin{center}
\begin{tabular}{|c|c|c|}
\hline
&&\\
One dimensional  & Fully connected & Fully connected\\
model&  model,&  model, \\
 on a ring & two reservoirs &one bath at $T=0$\\
&&\\
\hline
&&\\
$\Delta S^{1/2} \propto \frac{|\beta_p-\beta_0|}{\beta}$ & $\Delta S^{1/2} \propto \frac{|\beta_p-\beta_0|}{\beta}$ & $\Delta S= 0$, $\beta_p=\beta_0$ \\ 
&&\\
\hline  
&&\\
$\sigma_s=0$ & $\sigma_s \propto \zeta(\nu)\frac{|\beta_p-\beta_0|}{\beta}$ & $\sigma_s\to\infty$ \\ 
&&\\
\hline 
&&\\
observable  & observable  &  no observable\\ 
dependence & dependence & dependence\\ 
&&\\
\hline
\end{tabular}
\end{center}

\caption{Summary of the results obtained for the three different models
used as illustration, recalling the values of the entropy difference
$\Delta S$ and of the entropy production $\sigma_s$, together with
the observable-dependence of the FDR-temperature.}
\label{table}
\end{table}

It would be interesting to further test the present approach in experiments
or numerical simulations of realistic models. One possibility would be
to measure on the one hand the FDR-temperature for different observables
and different driving intensities, and on the other hand the
correction $F(\ve)$ to the equilibrium distribution, from which
$\Delta S$ could be evaluated. This independent determination of
$\beta_p$ and of $\Delta S$ would then allow for a test of the relation
(\ref{observable-entropy}) between these two quantities.
Alternatively, assuming the validity of Eq.~(\ref{observable-entropy}),
one could estimate the order of magnitude of the observable-dependence
of the FDR-temperature from the knowledge of $\Delta S$
(assuming that $\kappa_p$ is of order unity) or, in the
opposite way, assess $\Delta S$ from measurements of the temperatures
associated to different observables.

We argued that in out-of-equilibrium systems, the effective
temperature generically depends on the observable.
Our derivation relies on some rather strong assumptions
(statistical independence of the degrees of freedom, and local
decorrelation by each dynamical event), but there is no reason
to imagine that the observable-independence of the FDR-temperature
would be restored when such assumptions are not fulfilled.
We also believe that our arguments qualitatively extend beyond
the perturbative regime obtained for weak driving forces,
in the sense that we expect the non-uniformity of the phase-space distribution
on energy shells to yield, in a generic way, an observable-dependence
of the FDR-temperature (even though Eq.~(\ref{observable-entropy})
may not be valid in a strong forcing regime).
Hence the notion of effective temperature defined from
fluctuation-dissipation relations in non-equilibrium systems
seems to have a limited range of applicability.
Recently, others types of generalization have been proposed, not relying
on a notion of temperature, but rather relating the response function
to a suitable, and often more complicated, correlation function
\cite{Sasa,Villamaina,Maes,Prost,Gomez}.
Such an approach is certainly promising as it allows the deviations
from the equilibrium FDR to be understood in more details
\cite{Maes}. These deviations often appear in the form of an additive
term \cite{Sasa,Chetrite}, as can also be seen from Eqs.~(\ref{cp-general})
and (\ref{chip-general}).
In the framework of Langevin equations, such additive corrections
have been interpreted in terms of dissipated energy flux \cite{Sasa}.
In some case, for instance when a particle is trapped in a moving
potential well, an equilibrium form of the FDR can be restored by moving from
the standard eulerian frame to the lagrangian frame associated to the trap
\cite{Gomez}.
Finally, we note that it would be interesting to further clarify the link
between the present work and the results of \cite{Cugliandolo97}.
In the latter, an upper bound for the deviation from equilibrium FDR
was reported in the context of Langevin equations. This upper bound is
a function of the entropy production, and implies that there should be
no deviation from the equilibrium FDR if the entropy production is zero.
In the ring model presented in Sec.~\ref{sec-ring}, we found a systematic
deviation in the FDR, although the entropy production remains equal to zero.
Although there is strictly speaking no contradiction with the results of
\cite{Cugliandolo97} since the latter apply to Langevin systems,
it would be interesting to understand through which mechanism the bound
provided by \cite{Cugliandolo97} can be violated.

\subsection*{Acknowledgments}
This work has been partly supported by the Swiss National Science Foundation.

\appendix

\section{Calculation of $\Delta$S}
\label{app-deltaS}

In order to calculate the entropy difference $\Delta S = S_\mathrm{eq}(\beta^*) - S(\beta,\gamma)$ to leading order in the driving $\gamma$, we need to
include in the expansion (\ref{dist-dev}) the second order term in $\gamma$:
\bea \label{app-dist}
p(x,h) &=& p_\mathrm{eq}(x,h) \left[1+\gamma F(\ve_h(x))\right.\\ \nonumber
&& \qquad \qquad \qquad
\left. +\gamma^2 G(\ve_h(x)) + \mathcal{O}(\gamma^3)\right].
\eea
To lighten the notation we shall omit the index and the variable dependence of $\ve_h(x)$ in the following and simply write $\ve$ instead.
Let us first concentrate on the entropy of the nonequilibrium steady state in the presence of the forcing $\gamma$.
Expanding $S$ to order $\gamma^2$, we get
\begin{eqnarray} \label{Sbg}
S(\beta,\gamma)&=& S_\mathrm{eq}(\beta)+\gamma\beta\langle \varepsilon F(\varepsilon)\rangle_\mathrm{eq}\\
&+& \gamma^2\left[ \beta\langle \varepsilon G(\varepsilon)\rangle_\mathrm{eq}-\frac{1}{2} \langle F(\varepsilon)^2\rangle_\mathrm{eq}\right]
+\mathcal{O}(\gamma^3)\;,
\nonumber
\end{eqnarray}
where we took into account that the equilibrium averages
$\langle F(\ve)\rangle_\mathrm{eq}$ and $\langle G(\ve)\rangle_\mathrm{eq}$
are identically zero due to the normalization condition on $p(x,h)$.
We denote as $\ve_\mathrm{eq}(\beta)$ the equilibrium average energy
per degree of freedom, $\langle \ve\rangle_\mathrm{eq}$, at temperature
$\beta$. The nonequilibrium average energy $\langle \ve\rangle$
in the presence of a forcing $\gamma$ is denoted as
$\ve(\beta,\gamma)$.
To calculate the equilibrium entropy at temperature $\beta^*$, 
\begin{equation} \label{Seq-beta-star-app}
S_\mathrm{eq}(\beta^*)=\ln Z(\beta^*)+\beta^* \ve_\mathrm{eq}(\beta^*) \;,
\end{equation}
we write $\beta^*$ as a second order expansion in $\gamma$
\begin{equation}
\beta^*=\beta^*_0+\gamma \beta^*_1+\gamma^2 \beta^*_2+\mathcal{O}(\gamma^3)\;.
\end{equation}
Remember that the temperature $\beta^*$ is defined implicitly by $\ve_\mathrm{eq}(\beta^*)=\ve(\beta,\gamma)$. Thus we can determine the coefficients in the above expression by comparing the expansion of $\ve_\mathrm{eq}(\beta^*)$ with the expansion of $\ve(\beta,\gamma)$. 
This yields $\beta^*_0=\beta$ and further
\begin{eqnarray} \label{eq-beta1-star-app}
\ve'_\mathrm{eq}(\beta) \beta_1^* &=& \langle \varepsilon F(\varepsilon) \rangle_\mathrm{eq}\\
\ve'_\mathrm{eq}(\beta) \beta_2^*+\frac{1}{2} \ve''_\mathrm{eq}(\beta) \beta_1^{*2} &=& \langle \varepsilon G(\varepsilon) \rangle_\mathrm{eq}
\end{eqnarray}
where  $\ve'_\mathrm{eq}(\beta)$ and $\ve''_\mathrm{eq}(\beta)$ denote the first, respectively the second, derivative of $\ve_\mathrm{eq}(\beta)$.
Note that here and below, the equilibrium average
$\langle \ldots\rangle_\mathrm{eq}$ is evaluated at temperature $\beta$.

We now proceed to compute $S_\mathrm{eq}(\beta^*)$ from
Eq.~(\ref{Seq-beta-star-app}).
The expansion of $\ln Z(\beta^*)$ reads
\begin{eqnarray}
\ln Z(\beta^*)&=& \ln Z(\beta)-\gamma\beta^*_1 \langle\ve\rangle_\mathrm{eq}\nonumber\\
&&-\gamma^2\left[\beta^*_2\langle\ve\rangle_\mathrm{eq}+\frac{1}{2}\beta^*_1 \langle\ve F(\ve)\rangle_\mathrm{eq}\right]\nonumber\\
&&+\mathcal{O}(\gamma^3) \;.
\end{eqnarray}
From the expansion of $\ve(\beta,\gamma)$,
\be
\ve(\beta,\gamma) = \ve_\mathrm{eq}(\beta)
+\gamma \langle \ve F(\ve)\rangle_\mathrm{eq}
+\gamma^2 \langle \ve G(\ve)\rangle_\mathrm{eq} + \mathcal{O}(\gamma^3),
\ee
and the expansion of $\beta^*$, we obtain
\begin{eqnarray}
\beta^* \ve_\mathrm{eq}(\beta^*)&=& \beta^* \ve(\beta,\gamma)\nonumber\\
&=&\beta\langle\ve\rangle_\mathrm{eq}+\gamma\left[\beta^*_1 \langle\ve\rangle_\mathrm{eq}+\beta\langle\ve F(\ve)\rangle_\mathrm{eq}\right]\nonumber\\
&&+\gamma^2\left[\beta^*_2\langle\ve\rangle_\mathrm{eq}+\beta^*_1\langle\ve F(\ve)\rangle_\mathrm{eq}+\beta\langle\ve G(\ve)\rangle_\mathrm{eq}\right]\nonumber\\
&&+\mathcal{O}(\gamma^3)
\end{eqnarray}
so that we finally get for the equilibrium entropy at $\beta^*$:
\begin{eqnarray}
S_\mathrm{eq}(\beta^*) &=& S_\mathrm{eq}(\beta)+\gamma\beta\langle\varepsilon F(\varepsilon)\rangle_\mathrm{eq} \nonumber\\
&&+\gamma^2 \left[\beta\langle\varepsilon G(\varepsilon)\rangle_\mathrm{eq}+\frac{\beta_1^*}{2}\langle\varepsilon F(\varepsilon)\rangle_\mathrm{eq}\right]
\end{eqnarray}
Using this expression and the result (\ref{Sbg}) yields for the
entropy difference
\begin{equation} \label{delta-S-app}
\Delta S=\frac{\gamma^2}{2}\left[\langle F(\varepsilon)^2\rangle_\mathrm{eq}-\frac{\langle\varepsilon F(\varepsilon)\rangle_\mathrm{eq}^2}{\langle \varepsilon^2\rangle_\mathrm{eq}-\langle \varepsilon\rangle_\mathrm{eq}^2}\right] \;,
\end{equation}
where we also took into account Eq.~(\ref{eq-beta1-star-app})
as well as the relation
$\ve'_\mathrm{eq}(\beta)=-(\langle \varepsilon^2\rangle_\mathrm{eq}-\langle \varepsilon\rangle_\mathrm{eq}^2)$.

Equation (\ref{delta-S-app}) can be rewritten as
\begin{eqnarray}
\Delta S&=&\frac{\gamma^2}{2\langle( \varepsilon-\langle \varepsilon\rangle)^2\rangle}\times\\
&&\left[\langle( \varepsilon-\langle \varepsilon\rangle)^2\rangle\langle F(\varepsilon)^2\rangle-\langle(\varepsilon-\langle\varepsilon\rangle) F(\varepsilon)\rangle^2\right]\;,
\nonumber
\end{eqnarray}
where we omitted the index on the brackets, which all indicate averaging with respect to the equilibrium distribution.
From this expression, one can deduce that $\Delta S\geq 0$,
since the prefactor is strictly positive and the expression in the brackets
is greater or equal to zero due to the Cauchy-Schwarz inequality.

\section{FDR for the ring model}
\label{app-fdr}

We provide in this appendix further details on the derivation
of the FDR for the ring model studied in Sect.~\ref{sec-ring}.
The derivation of the expression for the response function reduces to the
calculation of the logarithmic derivative of the probability
distribution with respect to $h$. Starting from
\be
\mathcal{P}(\{x_i\},h) = \frac{1}{Z_N} \prod_{i=1}^N
\frac{|x_i-h|^{2\delta-1}}{2^{\delta-1}}
\exp\left(-\frac{(x_i-h)^2}{2T}\right) \;,
\ee
we get
\begin{eqnarray}
\left.\frac{\partial \ln \mathcal{P}(\{x_i\},h)}{\partial h}\right|_{h=0}
= &-&\left.\frac{\partial \ln Z_N}{\partial h}\right|_{h=0}\\ \nonumber
&+& \sum_{i=1}^N\left(\frac{x_i}{T}-\frac{2\delta-1}{x_i}\right).
\end{eqnarray}
The first term on the right hand side is identically zero,
since $Z_N$ is independent of $h$, as can be seen by the simple
change of variable $v_i=x_i-h$ in the integral defining $Z_N$.

Using these results in the expression for the response $\chi_p(t)$,
given in (\ref{response-rel}), of the observable $B_p(t)$ yields
\begin{eqnarray}
\nonumber
\chi_p(t) &=& \left\langle\left(\sum_{i=1}^N\left\{\frac{x_i(0)}{T}
-\frac{2\delta-1}{x_i(0)}\right\}\right) \right.\\
\label{resp}
&& \qquad \qquad \qquad  \times
\left.\left(\sum_{i=1}^N x_i^{2p+1}(t)\right)\right\rangle.
\end{eqnarray}
As the variables $x_i$ and $x_j$ are independent for $i\ne j$, we find
\be \label{resp2}
\chi_p(t) = \frac{N}{T}\left\langle x(0)x(t)^{2p+1}\right\rangle
- N(2\delta-1)\left\langle\frac{x(t)^{2p+1}}{x(0)} \right\rangle\;.
\ee

Due to the random sign change in the definition of the dynamics, each event decorrelates the involved variables $x_i$ from their previous values. As a consequence we can express the different averages involved in expression (\ref{resp}) in terms of the persistence probability $\Phi(t)$ as explained in
Sect.~\ref{sec-frame}, which leads to
\be
\langle x(0)^{-1}x(t)^{2p+1}\rangle = \langle x^{2p}\rangle\Phi(t)\;.
\ee
As for the correlation function $C_p(t)$, we find from Eq.~(\ref{cp-ring})
\be
C_p(t) = N\langle x^{2p+2}\rangle\Phi(t) \label{cp} \;,
\ee
where the expression for the correlation function (\ref{cp}) is the same
as in Eq.~(\ref{cp-general}) with the special choice $\psi(x)=x$,
given by the coupling to the field (\ref{h-coupling}).
Using these relations in Eq.~(\ref{resp2}) greatly simplifies the
expression, yielding
\begin{eqnarray}\label{chi}
\chi_p(t)=\left(\frac{1}{T}-\frac{(2\delta-1)\langle x^{2p}\rangle}
{\langle x^{2p+2}\rangle}\right)C_p(t)\;.
\end{eqnarray}

To obtain the final result, we have to calculate the even moments of $x$,
which can be easily done due to the complete factorization of the
probability distribution. The zero-field one-site distribution $p_0(x)$
is given by
\begin{equation}
p_0(x) = \frac{|x|^{2\delta-1}}{(2T)^{\delta} \Gamma(\delta)}\, e^{-x^2/2T}\;,
\end{equation}
with $\Gamma(\delta) = \int_0^\infty dz\,z^{\delta-1} e^{-z}$
being the Euler Gamma function. Using these results we can calculate
the moments as
\begin{eqnarray}
\langle x^{2n}\rangle&=&(2T)^{n}\frac{\Gamma(\delta+n)}{\Gamma(\delta)}\;,
\end{eqnarray}
for $n\geq 0$ integer, from which Eq.~(\ref{chip-ring}) follows,
using (\ref{chi}).

\section{Best polynomial approximation for $F(\ve)$}
\label{app-F}

In this appendix, we explain the approximation scheme allowing us to
derive the best polynomial approximation for $F(\ve)$.
The basic idea is to linearize the evolution equation for $P(\ve)$
in the parameter $\gamma=\lambda^2$.
The approximation is then found through a variational method,
by minimizing the error in the linearized evolution equation,
under the constraint of normalization and zero net flux
in the system.

As an illustration of the method, we use the simplest example
of a second order polynomial
$F^{(2)}(\ve)=a_0^{(2)}+a_1^{(2)}\beta\ve+a_2^{(2)}(\beta\ve)^2$.
The method however applies to polynomials of arbitrary order.
We have considered polynomial approximations up to order $L=5$.
To lighten the expressions we set $\beta=1$, without loss of generality.

To satisfy the normalization constraint
$\langle F^{(L)}(\ve)\rangle_\mathrm{eq}=0$ for $F^{(2)}(\ve)$,
we obtain $a_0^{(2)}=-\frac{1}{2}a_1^{(2)}-\frac{3}{4}a_2^{(2)}$.
The balance of fluxes implies that the average energy flowing out 
\begin{equation}
|\mathcal{J}_\mathrm{out}| = 2\nu \int_0^\infty d\ve\; P(\ve) \int_0^\ve d\mu
\, \varphi(\mu|\ve)\mu\;,
\end{equation}
and into the system 
\begin{equation}
|\mathcal{J}_\mathrm{in}|=\int_0^\infty d\mu\, (J_1(\mu)+J_2(\mu))\mu\;,
\end{equation}
should be equal in absolute value
$|\mathcal{J}_\mathrm{in}|=|\mathcal{J}_\mathrm{out}|$.
This condition yields $a_0^{(2)}=-\frac{3}{4}+\frac{5}{4} a_2^{(2)}$
and $a_1^{(2)}=\frac{3}{2}-4a_2^{(2)}$.

We first need to obtain the equation satisfied by $F(\ve)$.
To this aim, we linearize the evolution equation
(\ref{P-evolution}) for the probability distribution $P(\ve,t)$
with respect to the parameter $\gamma=\lambda^2$.
Denoting as $R(\ve)$ the linearized r.h.s.~of Eq.~(\ref{P-evolution}),
we find
\begin{eqnarray*}
R(\ve)&=&\int_0^\ve d\mu\; \mu^2 J_0(\mu)P_\mathrm{eq}(\ve-\mu)\\
&&+2\int_0^\ve d\mu J_0(\mu)P_\mathrm{eq}(\ve-\mu)F(\ve-\mu)\\
&&-\int_0^\infty d\mu \mu^2 J_0(\mu)P_\mathrm{eq}(\ve)\\
&&-2\int_0^\infty d\mu J_0(\mu)P_\mathrm{eq}(\ve)F(\ve)\\
&&+(2\nu+1)\int_0^\infty d\mu \varphi(\mu|\ve+\mu)P_\mathrm{eq}(\ve+\mu)F(\ve+\mu)\\
&&-(2\nu+1)\int_0^\ve d\mu \varphi(\mu|\ve)P_\mathrm{eq}(\ve)F(\ve)\\
&&+\int_0^\ve d\mu P_\mathrm{eq}(\ve-\mu)F(\ve-\mu)\int_\mu^\infty d\ve' \varphi(\mu|\ve')P_\mathrm{eq}(\ve')\\
&&+\int_0^\ve d\mu P_\mathrm{eq}(\ve-\mu)\int_\mu^\infty d\ve' \varphi(\mu|\ve')P_\mathrm{eq}(\ve')F(\ve')\\
&&-\int_0^\infty d\mu P_\mathrm{eq}(\ve)F(\ve)\int_\mu^\infty d\ve' \varphi(\mu|\ve')P_\mathrm{eq}(\ve')\\
&&-\int_0^\infty d\mu P_\mathrm{eq}(\ve)\int_\mu^\infty d\ve' \varphi(\mu|\ve')P_\mathrm{eq}(\ve')F(\ve') \;,
\end{eqnarray*}
with $J_0(\mu)=\nu\, e^{-\beta\mu}$.
In the stationary state this expression should be equal to zero, yielding
\begin{eqnarray*}
0&=& \frac{16}{15}\nu\ve^3+(2\nu+1)\ve^{\frac{1}{2}}\int_0^\ve d\mu\, \mu^{-\frac{1}{2}}F(\mu)\\
&&-(2\nu+1)F(\ve)+(2\nu+1)\int_0^\infty d\mu\, e^{-\mu}F(\ve+\mu)\\
&&-2\nu-2(2\nu+1)\ve F(\ve)\\
&&+\left(\frac{\ve}{\pi}\right)^{\frac{1}{2}}\int_0^\ve d\mu (\ve-\mu)^{-\frac{1}{2}}\int_0^\infty d\rho\,\rho^{-\frac{1}{2}}e^{-\rho}F(\rho+\mu)\\
&&-\frac{2}{\sqrt{\pi}}\int_0^\infty d\mu\, \mu^{\frac{1}{2}}e^{-\mu}F(\mu)
\end{eqnarray*}
As the exact solution for $F(\ve)$ is hard to obtain, we
replace $F(\ve)$ by its approximation $F^{(2)}(\ve)$ in the r.h.s~of
the last equation, yielding
\begin{eqnarray}
R^{(2)}(\ve)&=&-a_2^{(2)}+\nu-4a_2^{(2)}\nu+(2a_2^{(2)}+4a_2^{(2)}\nu)\ve\nonumber\\
&&+\frac{4}{3}(a_2^{(2)}-3\nu+8a_2^{(2)}\nu)\ve^2\nonumber\\
&&-\frac{8}{15}(a_2^{(2)}-2\nu+6a_2^{(2)}\nu)\ve^3 \;,
\end{eqnarray}
which is not equal to zero. In order to minimize the error, we use a
variational procedure. We first define a norm for the function $R^{(2)}(\ve)$
as
\be
||R^{(2)}|| \equiv \langle R^{(2)}(\ve)^2\rangle_\mathrm{eq}^{\frac{1}{2}}.
\ee
Then we look for the value of $a_2^{(2)}$ that minimizes the norm
$||R^{(2)}||$, namely
\begin{equation}
\frac{d}{d a_2^{(2)}} ||R^{(2)}|| = 0 \;.
\end{equation}
Solving this equation, we find as best approximation for the coefficient
$a_2^{(2)}=a_2^{(2)}(\nu)$:
\begin{equation}
a_2^{(2)}(\nu)=\frac{3\nu(7+37\nu)}{13+136\nu+358\nu^2}\;.
\end{equation}
Higher order approximations can be obtained through similar calculations.

\end{document}